%% file: main.tex
\documentclass{article}

\usepackage{pifont}
    \usepackage[preprint]{neurips_2021}

\usepackage{stmaryrd}
\usepackage{amsmath}
\usepackage[utf8]{inputenc} 
\usepackage[T1]{fontenc}    
\usepackage[colorlinks,citecolor=magenta]{hyperref}       
\usepackage{url}            
\usepackage{booktabs}       
\usepackage{amsfonts}       
\usepackage{nicefrac}       
\usepackage{microtype}      
\usepackage{xcolor}         
\usepackage{graphicx}
\usepackage{todonotes}
\usepackage{natbib}
\usepackage{wrapfig}
\usepackage{enumitem}

\title{Towards Denotational Semantics of AD for Higher-Order, Recursive, Probabilistic Languages}

\author{%
  Alexander K. Lew\\
  MIT\\
  \And
  Mathieu Huot \\
  Oxford \\
  \And
  Vikash K. Mansinghka\\
  MIT\\
}

\begin{document}

\maketitle

\begin{abstract}
Automatic differentiation (AD) aims to compute derivatives of user-defined functions, but in Turing-complete languages, this simple specification does not fully capture AD's behavior: AD sometimes disagrees with the true derivative of a differentiable program, and when AD is applied to non-differentiable or effectful programs, it is unclear what guarantees (if any) hold of the resulting code. 
 We study an expressive differentiable programming language, with piecewise-analytic primitives, higher-order functions, and general recursion. Our main result is that even in this general setting, a version of \citet{lee2020correctness}'s correctness theorem (originally proven for a first-order language {\it without} partiality or recursion) holds: all programs denote so-called \textit{$\omega$PAP functions}, and AD computes correct \textit{intensional derivatives} of them. \citet{mazza2021automatic}'s recent theorem, that AD disagrees with the true derivative of a differentiable recursive program at a measure-zero set of inputs, can be derived as a straightforward corollary of this fact. We also apply the framework to study probabilistic programs, and recover a recent result from \citet{mak2021densities} via a novel denotational argument.
\end{abstract}

\section{Introduction}

Automatic differentiation (AD) refers to a family of techniques for mechanically computing derivatives of user-defined functions.  When applied to straight-line programs built from differentiable primitives (without \texttt{if} statements, loops, and other control flow), AD has a straightforward justification using the chain rule. But in the presence of more expressive programming constructs, AD is harder to reason about: some programs encode partial or non-differentiable functions, and even when programs \textit{are} differentiable, AD can fail to compute their true derivatives at some inputs.


The aim of this work is to provide a useful mathematical model of (1) the class of partial functions that recursive, higher-order programs built from ``AD-friendly'' primitives can express, and (2) the guarantees AD makes when applied to such programs. Our model helps answer questions like:
\begin{itemize}[leftmargin=*]
    \item \textbf{If AD is applied to a recursive program, does the derivative halt on the same inputs as the original program?} We show that the answer is yes, and that restricted to this common domain of definition, AD computes an \textit{intensional derivative} of the input program~\citep{lee2020correctness}.
    
    \item \textbf{Is it sound to use AD for ``Jacobian determinant'' corrections in probabilistic programming languages (PPLs)?} Many probabilistic programming systems use AD to compute densities of transformed random variables~\citep{radul2021base} and reversible-jump MCMC acceptance probabilities~\citep{cusumano2020automating}. AD produces correct derivatives for almost all inputs~\citep{mazza2021automatic}, but PPLs evaluate derivatives at inputs sampled by probabilistic programs, \textit{which may have support entirely on Lebesgue-measure-zero manifolds of $\mathbb{R}^n$}. We may wonder: could PPLs, for certain input programs, produce wrong answers \textit{with probability 1}? We show that fortunately, even when AD gives wrong answers, it does so \textit{in a way} that does not compromise the correctness of the Jacobian determinants that PPLs compute.
    %
\end{itemize}
Our approach is inspired by \citet{lee2020correctness}, who provided a similar characterization of AD for a first-order language with branching (but not recursion). Section 2 reviews their development, and gives intuition for why recursion, higher-order functions, and partiality present additional challenges. In Section~3, we give our solution to these challenges, culminating in a similar result to that of \citet{lee2020correctness} but for a more expressive language. We recover as a straightforward corollary the result of \citet{mazza2021automatic} that when applied to recursive, higher-order programs, AD fails at a measure-zero set of inputs. In Section 4, we briefly discuss the implications of our characterization for PPLs. Indeed, reasoning about AD in probabilistic programs is a key motivation for our work: perhaps more so than typical differentiable programs, probabilistic programs often employ higher-order functions and recursion as modeling tools. For example, early Church programs used recursive, higher-order functions to express non-parametric probabilistic grammars~\citep{goodman2012church}, and modern PPLs such as Gen~\citep{cusumano2019gen} and Pyro~\citep{bingham2019pyro} use higher-order primitives with custom derivatives or other specialized inference logic to scale to larger datasets.%
\footnote{Higher-order recursive combinators like \texttt{map} and \texttt{unfold} enforce conditional independence patterns that systems can exploit for subsampling-based gradient estimates (in Pyro) or incremental computation (in Gen).} Furthermore, as mentioned above, probabilistic programs may have support entirely on Lebesgue-measure-zero manifolds, so the intuition that AD is correct ``almost everywhere'' becomes less useful as a reasoning aide\textemdash motivating the need for more precise models of AD's behavior. 


{\bf Related work.} The growing importance of AD for learning and inference has inspired a torrent of work on the semantics of differentiable languages, summarized in Table~\ref{table:comparison}. We build on existing denotational approaches, particularly those of \citet{huot2020correctness} and \citet{vakar2020denotational}, but incorporate ideas from \citet{lee2020correctness} to handle piecewise functions, and \citet{vakar2019domain} to model probabilistic programs. \citet{mazza2021automatic} consider a language equally expressive as our deterministic fragment; we give a novel and complementary denotational account (their approach is operational). \citet{mak2021densities} do not consider AD, but do give an operational semantics for a Turing-complete PPL, and tools for reasoning about differentiability of density functions. 

\begin{table}[]
\centering
\footnotesize{
\begin{tabular}{|r|c|c|c|c|c|c|}
\hline
{\bf Semantic Framework} & {\bf Piecewise} & {\bf Recursion}  & {\bf Higher-Order} & {\bf Approach} & {\bf AD} & {\bf PPL} \\\hline
\citet{huot2020correctness} &  &  & \ding{52} & Denotational & \ding{52} &  \\
\citet{vakar2020denotational} &  & \ding{52} & \ding{52} & Denotational & \ding{52} &  \\
\citet{lee2020correctness} & \ding{52} &  &  & Denotational & \ding{52} &  \\
\citet{abadi-plotkin2020} &  & \ding{52} &  & Both & \ding{52} &  \\
\citet{mazza2021automatic} & \ding{52} & \ding{52} & \ding{52} & Operational & \ding{52} &  \\
\citet{mak2021densities} & \ding{52} & \ding{52} & \ding{52} & Operational &  & \ding{52} \\
Ours & \ding{52} & \ding{52} & \ding{52} & Denotational & \ding{52} & \ding{52}\\\hline
\end{tabular}
}
\vspace{1mm}
\caption{Approaches to reasoning about differentiable programming languages and AD. ``Piecewise'': the semantics accounts for total but discontinuous functions such as $<$ and $==$ for reals. ``PPL'': the same semantic framework can be used to reason about probabilistic programs and the differentiable properties of deterministic ones. ``AD'': the framework supports reasoning about soundness of AD (NB: we handle only forward-mode, whereas others handle reverse-mode or both).}
\vspace{-8mm}
\label{table:comparison}
\end{table}

\vspace{-3mm}
\section{Background: PAP functions and intensional derivatives}
\vspace{-2mm}
In this section, we recall \citet{lee2020correctness}'s approach to understanding a simple differentiable programming language, and describe the key challenges for extending their approach to a more complex language, with partiality, recursion, and higher-order functions.
\vspace{-3mm}

\subsection{A First-Order Differentiable Programming Language}
\vspace{-2mm}
\citet{lee2020correctness} consider a first-order language with real number constants $c$, primitive real-valued functions $f : \mathbb{R}^{N_f} \rightarrow \mathbb{R}$, as well as an \texttt{if} construct for branching: $e ::= c \mid x_i \mid f(e_1,...,e_n) \mid \texttt{if} \, (e_1 > 0) \, e_2 \, e_3$. Expressions $e$ in the language denote functions $\llbracket e \rrbracket : \mathbb{R}^n \rightarrow \mathbb{R}$ of an \textit{input vector} $\mathbf{x} \in \mathbb{R}^n$ (for some $n$). We have $\llbracket c\rrbracket \mathbf{x} = c$, $\llbracket x_i \rrbracket \mathbf{x} = \mathbf{x}[i]$, $\llbracket f(e_1, \dots, e_n) \rrbracket \mathbf{x} = f(\llbracket e_1\rrbracket \mathbf{x}, \dots, \llbracket e_n\rrbracket\mathbf{x})$, and $\llbracket \texttt{if }(e_1 > 0)\, e_2\, e_3\rrbracket \mathbf{x} = [ \llbracket e_1\rrbracket\mathbf{x} > 0] \cdot (\llbracket e_2\rrbracket \mathbf{x}) + [\llbracket e_1\rrbracket\mathbf{x} \leq 0] \cdot (\llbracket e_3\rrbracket\mathbf{x})$.

A key insight of \citet{lee2020correctness} is that if the primitive functions $f$ are \textit{piecewise analytic under analytic partition}, or \textit{PAP}, then so is any program written in the language. They define PAP functions in stages, starting with the concept of a \textit{piecewise representation} of a function $f$:

\textbf{Definition.} Let $U \subseteq \mathbb{R}^n$, $V \subseteq \mathbb{R}^m$, and $f : U \rightarrow V$. A \textit{piecewise representation} of $f$ is a countable family $\{(A_i, f_i)\}_{i \in I}$ such that: (1) the sets $A_i$ form a partition of $U$; (2) each ${f_i : U_i \rightarrow \mathbb{R}^m}$ is defined on an open domain $U_i\supseteq A_i$; and (3) when $x \in A_i$, $f_i(x) = f(x)$.

\begin{wrapfigure}{r}{0.55\textwidth}
    \centering
    \includegraphics[width=0.5\textwidth]{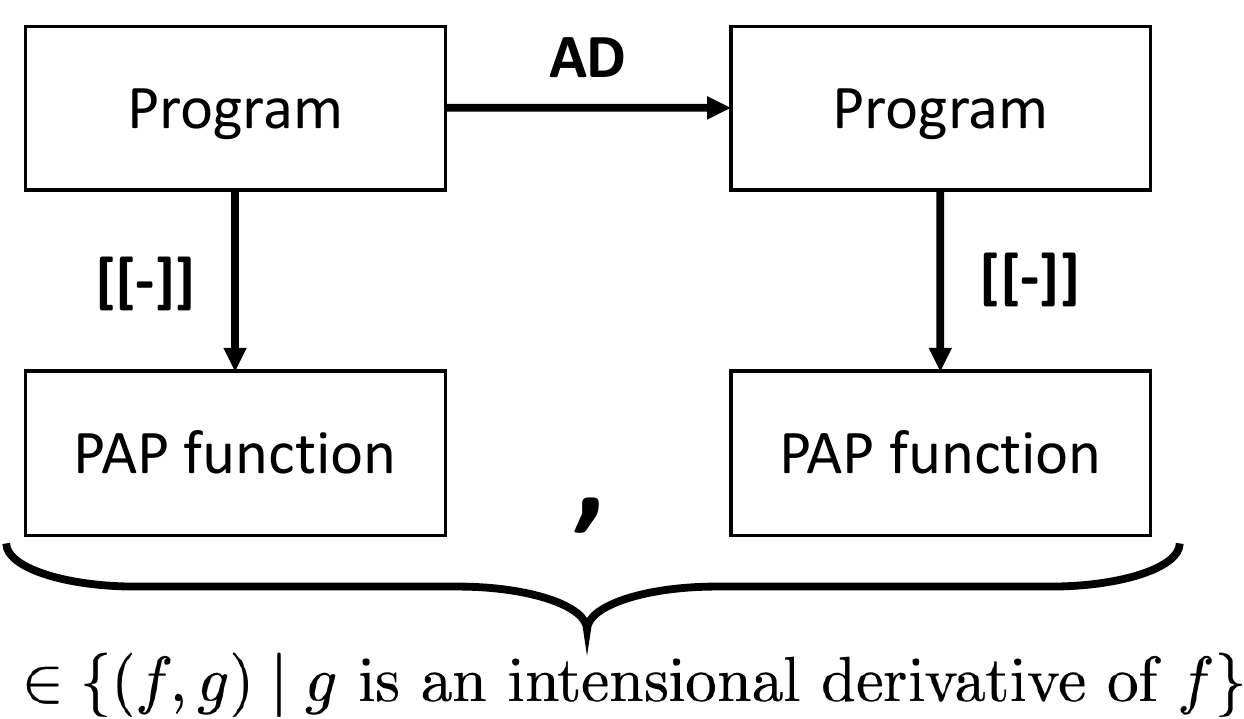}
    \caption{\citet{lee2020correctness}'s denotational characterization of AD, which we extend. For a simple differentiable programming language, they define a class of functions (\textit{PAP functions}) expressive enough to interpret all programs in the language. Not all PAP functions are differentiable, but \citet{lee2020correctness} propose a relaxed notion of derivatives, called \textit{intensional derivatives}, that always exist for PAP functions, and show that AD yields intensional derivatives. We extend their argument to handle recursion and higher-order functions, by defining a \textit{generalized} class of \textit{$\omega$PAP functions} and a corresponding generalization of the notion of intensional derivative.}
    \vspace{-10mm}
    \label{fig:my_label}
\end{wrapfigure}

The PAP functions are those with \textit{analytic} piecewise representations:

\textbf{Definition.} If the Taylor series of a smooth function $f$ converges pointwise to $f$ in a neighborhood around $x$, we say $f$ is \textit{analytic} at $x$. An \textit{analytic function} is analytic everywhere in its domain.
We call a set $A \subseteq \mathbb{R}^n$ an \textit{analytic set} iff there exist finite collections $\{g^+_i\}_{i \in I}$ and $\{g^-_j\}_{j \in J}$ of analytic functions into $\mathbb{R}$, with open domains $X^+_i$ and $X^-_j$, such that $A = \{x \in (\bigcap_{i} X^+_i) \cap (\bigcap_{j} X^-_j) \mid \forall i. g^+_i(x) > 0 \wedge \forall j. g^-_j \leq 0\}$ (i.e., analytic sets are subsets of open sets carved out using a finite number of analytic inequalities.)

\textbf{Definition.} We say $f$ is \textit{piecewise analytic under analytic partition (PAP)} if there exists a piecewise representation $\{(A_i, f_i)\}_{i \in I}$ of $f$ such that the $A_i$ are analytic sets and the $f_i$ are analytic functions. We call such a representation a \textit{PAP representation}.

\textbf{Proposition (Lee et al. 2020).} Constant functions and projection functions are PAP. Supposing $f : \mathbb{R}^n \rightarrow \mathbb{R}$ is PAP, and $\llbracket e_1\rrbracket, \dots, \llbracket e_n\rrbracket$ are all PAP, $\llbracket f(e_1, \dots, e_n)\rrbracket$ and $\llbracket \texttt{if } (e_1 > 0) \, e_2 \, e_3\rrbracket$ are also PAP. Therefore, by induction, all expressions $e$ denote PAP functions $\llbracket e\rrbracket$. 

PAP functions are not necessarily differentiable. But \citet{lee2020correctness} define \textit{intensional derivatives}, which do always exist for PAP functions (though unlike traditional derivatives, they are not unique):

\textbf{Definition.} A function $g$ is \textit{an intensional derivative} of a PAP function $f$ if there exists a PAP representation $\{(A_i, f_i)\}_{i \in I}$ of $f$ such that when $x \in A_i$, $g(x) = f_i'(x)$.

\citet{lee2020correctness} then give a standard AD algorithm for their language and show that, when applied to an expression $e$, it is guaranteed to yield \textit{some} intensional derivative of $\llbracket e\rrbracket$ as long as each primitive $f$ comes equipped with an intensional derivative. Essentially, this proof is based on an analogue to the chain rule for intensional derivatives. The result is depicted schematically in Figure~\ref{fig:my_label}.



\vspace{-2mm}
\subsection{Challenges: Partiality, Higher-Order Functions, and Recursion}
\begin{wrapfigure}{l}{0.40\textwidth}
\vspace{-6mm}
    \centering
    \includegraphics[width=0.35\textwidth]{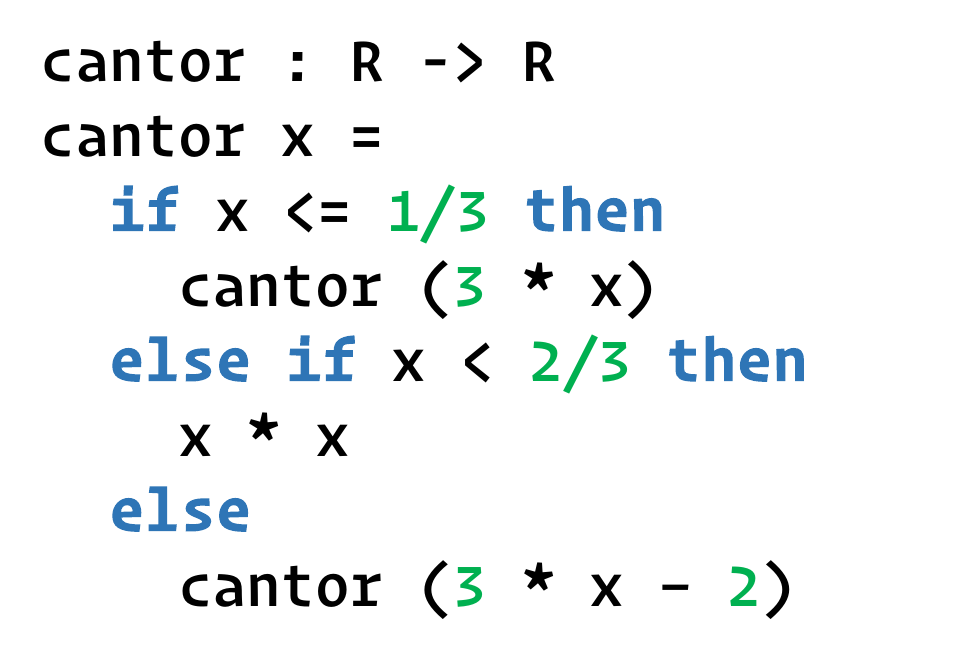}
    \caption{This program can be seen as a piecewise function, constantly $\bot$ on the $\frac{1}{3}$-Cantor set and $x^2$ elsewhere in $(0, 1)$. But the $\frac{1}{3}$-Cantor set is not expressible as a countable union of analytic sets, and so this representation is not PAP.}
    \vspace{-4mm}
    \label{fig:cantor_func}
\end{wrapfigure} 
Can a similar analysis be carried out for a more complex language, with higher-order and recursive functions? One challenge is that as defined above, only total, first-order functions can be PAP: unless we can generalize the definition to cover partial and higher-order functions, we cannot reproduce the inductive proof that PAP functions are closed under the programming language's constructs. This is a roadblock even if we care only about differentiating first-order, total programs. 
To see why, recall that we required primitives $f$ to be PAP in the section above.
What alternative requirements should we place on partial or higher-order primitives, to ensure that first-order programs built from them will be PAP? How should these primitives' built-in intensional derivatives behave?
    
    


These are non-trivial challenges, and it is possible to formulate reasonable-sounding solutions that turn out not to work. For example, we might hypothesize that \textit{partial} functions $f : U \rightharpoonup V$ definable using recursion are \textit{almost} PAP: perhaps there still exists an analytic partition $\{A_i\}_{i \in I}$ of $U$, and analytic functions $\{f_j\}_{j \in J}$ for some \textit{subset} of indices $J \subseteq I$, such that $f$ is defined exactly on $\bigcup_{j \in J} A_j$ and $f(x) = f_j(x)$ whenever $x \in A_j$. But consider the program $\texttt{cantor}$ in Figure 2. It denotes a partial function that is undefined outside of $[0, 1]$, and also on the $\frac{1}{3}$-Cantor set. This region \textit{cannot} be expressed as a countable union of analytic sets $\cup_{i \in I \setminus J} A_i$. Therefore, this candidate notion of PAP partial function is too restrictive: some recursive programs do not satisfy it.



\section{$\omega$PAP Semantics}
\vspace{-3mm}
In this section, we present an expressive differentiable programming language, with higher-order functions, branching, and general recursion. We then generalize the definitions of PAP functions and intensional derivatives to include higher-order and partial functions, and show that: (1) all programs in our language denote (this generalized variant of) PAP functions, and (2) a standard forward-mode AD algorithm computes valid intensional derivatives.

\vspace{-2mm}
\subsection{A Higher-Order, Recursive Differentiable Language}

\textbf{Syntax.} Consider a language with types $\tau ::= \mathbb{R}^k \mid \tau_1 \times \tau_2 \mid \tau_1 \rightarrow \tau_2$, and terms $e ::= x \mid c \mid e_1\, e_2 \mid \texttt{if } (e_1 > 0) \, e_2\, e_3 \mid \lambda x : \tau. e \mid \mu f : \tau_1 \rightarrow \tau_2. e$. Here, $c$ ranges over constants \textit{of any type}, including constant numeric literals such as $\texttt{3}$ and $\pi$, as well as primitive functions such as $\texttt{+}$ and $\texttt{sin}$. We write $(e_1, e_2)$ as sugar for $\texttt{pair}_{\tau_1, \tau_2}\, e_1\, e_2$, where $\texttt{pair}_{\tau_1, \tau_2} : \tau_1 \rightarrow \tau_2 \rightarrow \tau_1 \times \tau_2$ is a constant for constructing tuples. The $\mu$ expression creates a recursive function of type $\tau_1 \rightarrow \tau_2$, binding the name $f$ for the recursive call. For example, a version of the factorial function that works on natural number inputs is $\mu f : \mathbb{R} \rightarrow \mathbb{R}. (\lambda x : \mathbb{R}. \texttt{if } (x > 0) \,\, (x * f (x-1)) \,\, 1)$.

\textbf{Semantics of types.} For each type $\tau$ we choose a set $\llbracket \tau\rrbracket$ of values. We have $\llbracket\mathbb{R}^k\rrbracket = \mathbb{R}^k$ and $\llbracket\tau_1 \times \tau_2\rrbracket = \llbracket\tau_1\rrbracket \times \llbracket\tau_2\rrbracket$. Function types are slightly more complicated because we wish to represent \textit{partial} functions. Given a set $X$, we define $X_\bot = \{\uparrow x \mid x \in X\} \cup \{\bot\}$. (The tag $\uparrow$ is useful to avoid ambiguity when $\bot$ is already a member of $X$: then $X_\bot$ contains as distinct elements the newly adjoined $\bot$ and the representation $\uparrow \bot$ of the original $\bot$ from $X$.) Using this construction, we define $\llbracket \tau_1 \rightarrow \tau_2 \rrbracket = \llbracket\tau_1\rrbracket \rightarrow \llbracket\tau_2\rrbracket_\bot$: we represent partial functions returning $\tau_2$ as total functions into $\llbracket\tau_2\rrbracket_\bot$.

\textbf{Semantics of terms.} We interpret expressions of type $\tau$ as functions from \textit{environments} $\gamma$ (mapping variables $x$ to their values $\gamma[x]$) to $\llbracket\tau\rrbracket_\bot$. If $a \in \llbracket\tau\rrbracket_\bot$ and $b \in \llbracket\tau \rightarrow \sigma\rrbracket$, we write (as a mathematical expression, not an object language expression) $a \texttt{>>=} b$ to mean $\bot$ if $a = \bot$, and $b(x)$ if $a = \, \uparrow x$. Using this notation, we can define the interpretation of each construct in our language. We define: $\llbracket c\rrbracket\gamma = \uparrow c$, $\llbracket x\rrbracket\gamma = \uparrow(\gamma[x])$, $\llbracket e_1\, e_2\rrbracket\gamma = (\llbracket e_1\rrbracket\gamma) \texttt{>>=} (\lambda f. \llbracket e_2\rrbracket\gamma \texttt{>>=} f)$, $\llbracket \lambda x. e\rrbracket\gamma = \uparrow \lambda v. \llbracket e\rrbracket(\gamma[x \mapsto v])$, and $\llbracket \texttt{if } (e_1 > 0) \, e_2 \, e_3\rrbracket\gamma = \llbracket e_1\rrbracket\gamma \texttt{>>=} \lambda x. [x > 0 \mapsto \llbracket e_2\rrbracket\gamma, x \leq 0 \mapsto \llbracket e_3\rrbracket\gamma]$. To interpret recursion, we use the standard domain-theoretic approach. We first define partial orders $\preceq_\tau$ inductively for each type $\tau$: we define $x \preceq_{\mathbb{R}^k} y \iff x = y$, $(x_1, x_2) \preceq_{\tau_1 \times \tau_2} (y_1, y_2) \iff x_1 \preceq_{\tau_1} y_1 \wedge x_2 \preceq_{\tau_2} y_2$, and $f \preceq_{\tau \rightarrow \sigma} g \iff \forall x \in \llbracket\tau\rrbracket, f(x) \preceq^\bot_\sigma g(x)$. (The relation $x \preceq^\bot_\tau y$ holds when $x = \bot$ or when $x = \uparrow a$, $y = \uparrow b$, and $a \preceq_\tau b$ for some $a, b \in \llbracket\tau\rrbracket$.) Intuitively, $x \preceq y$ if $y$ is ``at least as defined'' as $x$. For the types $\tau$ in our language, given an infinite non-decreasing sequence $x_1 \preceq_\tau x_2 \preceq_\tau \dots$ of values, there exists a least upper bound $\vee_{i\in\mathbb{N}} x_i$. If the primitives in our language are \textit{Scott-continuous} (monotone with respect to $\preceq$, with the property that $\bigvee_i f(x_i) = f(\bigvee_i x_i)$), we can interpret recursion: $\llbracket \mu f : \tau \rightarrow \sigma. \lambda x. e\rrbracket\gamma = \uparrow \bigvee_{i \in \mathbb{N}} f_i$, where $f_0 = \lambda x. \bot$ and $f_i = \lambda v. \llbracket e\rrbracket(\gamma[x \mapsto v, f \mapsto f_{i-1}])$.


\subsection{$\omega$PAP functions}
We have given a standard denotational semantics to our language, interpreting terms as partial functions. We now generalize the notion of a PAP function to that of an \textit{$\omega$PAP partial function}, and show that if all the primitives are $\omega$PAP, so is any program. The definition relies on the choice, for each type $\tau$, of a set of well-behaved or ``PAP-like'' functions from Euclidean space \textit{into} $\llbracket\tau\rrbracket$, called the \textit{$\omega$PAP diffeology} of $\tau$, by analogy with diffeological spaces~\citep{iglesias2013diffeology}. 

\textbf{Definition.}  A set $U \subseteq \mathbb{R}^n$ is \textit{c-analytic} if it is a countable union of analytic sets.

\textbf{Definition.} Let $\tau$ be a type. An \textit{$\omega$PAP diffeology} $\mathcal{P}_\tau$ for $\tau$ assigns to each c-analytic set $U$ a set $\mathcal{P}^U_\tau$ of \textit{PAP plots in $\tau$}, functions from $U$ into $\llbracket\tau\rrbracket$ satisfying the following closure properties:
\vspace{-2mm}
\begin{itemize}
    \item \textbf{(Constants.)} All constant functions are plots in $\tau$.
    \item \textbf{(Closure under PAP precomposition.)} If $V \subseteq \mathbb{R}^m$ is a c-analytic set, and $f : V \rightarrow U$ is PAP, then $\phi \circ f$ is a plot in $\tau$ if $\phi : U \rightarrow \llbracket\tau\rrbracket$ is.
    \item \textbf{(Closure under piecewise gluing.)} If $\phi : U \rightarrow \llbracket\tau\rrbracket$ is such that the restriction $\phi|_{A_i} : A_i \rightarrow \llbracket\tau\rrbracket$ is a plot in $\tau$ for each $A_i$ in some c-analytic partition of $U$, then $\phi$ is a plot in $\tau$.
    \item \textbf{(Closure under least upper bounds.)} Suppose $\phi_1, \phi_2, \dots$ is a sequence of plots in $\tau$ such that for all $x \in U$, $\phi_i(x) \preceq_\tau \phi_{i+1}(x)$. Then $\bigvee_{i \in \mathbb{N}} \phi_i = \lambda x. \bigvee_{i \in \mathbb{N}} \phi_i(x)$ is a plot. 
\end{itemize}
\vspace{-3mm}
\textbf{Choosing $\omega$PAP diffeologies.} We set $\mathcal{P}^U_{\mathbb{R}^k}$ to be all PAP functions from $U$ to $\mathbb{R}^k$. (These trivially satisfy condition 4 above, since $\preceq_{\mathbb{R}^k}$ only relates equal vectors.) For $\tau_1 \times \tau_2$, we include $f \in \mathcal{P}^U_{\tau_1 \times \tau_2}$ iff $f \circ \pi_1 \in \mathcal{P}^U_{\tau_1}$ and $f \circ \pi_2 \in \mathcal{P}^U_{\tau_2}$. Function types are more interesting. A function $f : U \rightarrow \llbracket \tau_1 \rightarrow \tau_2\rrbracket$ is a plot if, for all PAP functions $\phi_1 : V \rightarrow U$ and plots $\phi_2 \in \mathcal{P}^V_{\tau_1}$, the function $\lambda v. f(\phi_1(v))(\phi_2(v))$ is defined (i.e., not $\bot$) on a c-analytic subset of $V$, restricted to which it is a plot in $\tau_2$.

\textbf{Definition.} Let $\tau_1$ and $\tau_2$ be types. Then a \textit{partial $\omega$PAP function} $f : \tau_1 \rightarrow \tau_2$ is a Scott-continuous function from $\llbracket\tau_1\rrbracket$ to $\llbracket\tau_2\rrbracket_\bot$, such that if $\phi \in \mathcal{P}^U_{\tau_1}$, $f \circ \phi$ is defined on a c-analytic subset of $U$, restricted to which it is a plot in $\tau_2$.%
\footnote{For readers familiar with category theory, the $\omega$PAP spaces ($\omega$cpos $(X, \preceq_X)$ equipped with $\omega$PAP diffeologies $\mathcal{P}_X$) and the total $\omega$PAP functions form a CCC, enriched over $\omega$Cpo. In fact, $\omega$PAP is equivalent to a category of models of an essentially algebraic theory, which means it also has all small limits and colimits. It can also be seen as a category of concrete sheaves valued in $\omega$Cpo, making it a Grothendieck quasi-topos.}
We revise our earlier interprtation of $\llbracket\tau_1 \rightarrow\tau_2\rrbracket$ to include only the partial $\omega$PAP functions.

We note that under this definition, a total function $f : \mathbb{R}^n \rightarrow \mathbb{R}^m$ is $\omega$PAP if and only if it is PAP. The generalization becomes apparent only when working with function types and partiality. 

\textbf{Proposition.} If every primitive function is $\omega$PAP, then every expression $e$ of type $\tau$ with free variables of type $\tau_1, \dots, \tau_n$ denotes a partial $\omega$PAP function $f : \tau_1 \times \dots \times \tau_n \rightarrow \tau$. In particular, programs that denote total functions from $\mathbb{R}^n$ to $\mathbb{R}^m$, even if they use recursion and higher-order functions internally, always denote (ordinary) PAP functions.

\vspace{-3mm}
\subsection{Automatic Differentiation}
\vspace{-3mm}
\textbf{Implementation of AD.} We now describe a standard forward-mode AD macro, adapted from \citet{huot2020correctness} and \citet{vakar2020denotational}. For each type $\tau$, we define a ``dual number type'' $\mathcal{D}\llbracket\tau\rrbracket$: $\mathcal{D}\llbracket\mathbb{R}^k\rrbracket = \mathbb{R}^k \times \mathbb{R}^k$, $\mathcal{D}\llbracket\tau_1 \times \tau_2\rrbracket = \mathcal{D}\llbracket\tau_1\rrbracket \times \mathcal{D}\llbracket\tau_2\rrbracket$, and $\mathcal{D}\llbracket\tau_1 \rightarrow \tau_2\rrbracket = \mathcal{D}\llbracket\tau_1\rrbracket \rightarrow \mathcal{D}\llbracket\tau_2\rrbracket$. 
The AD macro translates terms of type $\tau$ into terms of type $\mathcal{D}\llbracket\tau\rrbracket$: $\mathcal{D}\llbracket x\rrbracket = x$, $\mathcal{D}\llbracket e_1\, e_2\rrbracket = \mathcal{D}\llbracket e_1\rrbracket\, \, \mathcal{D}\llbracket e_2\rrbracket$, $\mathcal{D}\llbracket \texttt{if } (e_1 > 0) \, e_2\, e_3\rrbracket = \texttt{if } (\pi_1(\mathcal{D}\llbracket e_1\rrbracket) > 0)\,\,\mathcal{D}\llbracket e_2\rrbracket \,\, \mathcal{D}\llbracket e_3\rrbracket$, $\mathcal{D}\llbracket \lambda x: \tau. e\rrbracket = \lambda x: \mathcal{D}\llbracket\tau\rrbracket. \mathcal{D}\llbracket e\rrbracket$, and $\mathcal{D}\llbracket \mu f : \tau \rightarrow \sigma. e\rrbracket = \mu f : \mathcal{D} \llbracket\tau\rrbracket \rightarrow \mathcal{D}\llbracket\sigma\rrbracket. \mathcal{D}\llbracket e\rrbracket$. Constants $c$ come equipped with their own translations $c_\mathcal{D}$: $\mathcal{D}\llbracket c\rrbracket = c_\mathcal{D}$.
For constants of type $\mathbb{R}^k$, $c_\mathcal{D} = (c, \mathbf{0})$, but for functions, $c_\mathcal{D}$ encodes a primitive's intensional derivative. For example, when $c : \mathbb{R} \rightarrow \mathbb{R}$, we require that $c_\mathcal{D} : \mathbb{R} \times \mathbb{R} \rightarrow \mathbb{R} \times \mathbb{R}$ be such that, for any PAP function $f : \mathbb{R} \rightarrow \mathbb{R}$ with intensional derivative $g$, there is an intensional derivative $h$ of $c \circ f$ such that $c_\mathcal{D} \circ \langle f, g\rangle = \lambda x. ((c \circ f)(x), h(x))$. The constant \texttt{log}, for example, could have $\texttt{log}_\mathcal{D}((x, v)) = (\log x, \frac{v}{x})$.  



\textbf{Behavior / Correctness of AD.} For each type $\tau$, we define a family of relations $S^U_\tau \subseteq \mathcal{P}^U_\tau \times \mathcal{P}^U_{\mathcal{D}\llbracket\tau\rrbracket}$, indexed by the c-analytic sets $U$. The basic idea is that if $(f, g) \in S^U_\tau$, then $g$ is a correct ``intensional dual number representation'' of $f$. Since $S^U_\tau$ is a relation, there may be multiple such $g$'s, just as how in \citet{lee2020correctness}'s definition, a single PAP function may have multiple valid intensional derivatives.  

For $\tau = \mathbb{R}^k$, we build directly on \citet{lee2020correctness}'s notion of intensional derivative: $f$ and $g$ are related in $S^U_{\mathbb{R}^k}$ if and only if, for all PAP functions $h : V \rightarrow U$ and intensional derivatives $h'$ of $h$, there is some intensional derivative $G$ of $f \circ h$ such that for all $v \in V$, $g((h(v), h'(v))) = ((f \circ h)(v), G(v))$. For other types, we use the Artin gluing approach of \citet{vakar2020denotational} to derive canonical definitions of $S_\tau$ for products and function types that make the following result possible:

\textbf{Proposition.} Suppose $(\phi, \phi') \in S^U_\tau$, and that $e : \sigma$ has a single free variable $x : \tau$. Then $(\llbracket e\rrbracket \circ \phi, \llbracket\mathcal{D}\llbracket e\rrbracket\rrbracket \circ \phi')$ is defined on the same c-analytic subset $V \subseteq U$, and restricted to $V$, is in $S^V_\sigma$. 

Specializing to the case where $\tau$ and $\sigma$ are $\mathbb{R}^n$ and $\mathbb{R}^m$, this implies that AD gives sound intensional derivatives even when programs use recursion and higher-order functions. Intensional derivatives agree with derivatives almost everywhere, so this implies the result of~\citet{mazza2021automatic}.


\vspace{-4mm}
\section{Applications to Probabilistic Programming}
\label{sec:probabilistic}
\vspace{-3mm}
In this section, we briefly describe some applications of the $\omega$PAP semantics to \textit{probabilistic} languages. First, we give an example of applying our framework to reason about soundness for AD-powered PPL features. Second, we recover a recent result of~\citet{mak2021densities} via a novel denotational argument.

\textbf{AD for sound change-of-variables corrections.} Consider a PPL that represents primitive distributions by a pair of a \textit{sampler} $\mu$ and a \textit{density} $\rho$. Some systems support automatically generating a new sampler and density, $f_*\mu$ and $\rho_f$, for the \textit{pushforward} of $\mu$ by a user-specified deterministic bijection $f$. Such systems compute the density $\rho_f$ using a change-of-variables formula, which relies on $f$'s Jacobian~\citep{radul2021base}. We show in Appendix~\ref{sec:cov} that such algorithms are sound even when: (1) $f$ is not differentiable, but rather PAP; and (2) we use not $f$'s Jacobian but \textit{any} intensional Jacobian of $f$. This may be surprising, because intensional Jacobians can disagree with true Jacobians on Lebesgue-measure-zero sets of inputs, and the support of $\mu$ may lie entirely within such a set. Indeed, there are samplers $\mu$ and programs $f$ for which AD's Jacobians are wrong everywhere within $\mu$'s support. Our result shows that this does not matter: intensional Jacobians are ``wrong in the right ways,'' yielding correct densities $\rho_f$ even when the derivatives themselves are incorrect.


\textbf{Trace densities of probabilistic programs are $\omega$PAP.} Now consider extending our language with constructs for probabilistic programming: a type $\mathcal{M}\,\tau$ of $\tau$-valued probabilistic programs, and constructs $\mathbf{return}_\tau : \tau \rightarrow \mathcal{M}\,\tau$, $\mathbf{sample} : \mathcal{M} \mathbb{R}$, $\mathbf{score} : [0, \infty) \rightarrow \mathcal{M}\,\mathbf{1}$, and $\mathbf{do} \, \{x \gets t; s\} : \mathcal{M}\, \sigma$ (where $t : \mathcal{M}\,\tau$ and $s : \mathcal{M}\,\sigma$ in environments with $x : \tau$). Some PPLs use probabilistic programs only to specify density functions, for downstream use by inference algorithms like Hamiltonian Monte Carlo. To reason about such languages, we can interpret $\mathcal{M}\,\tau$ as comprising \textit{deterministic} functions computing values and densities of traces. More precisely, let $\llbracket\mathcal{M}\, \tau\rrbracket = \llbracket \sqcup_{i \in \mathbb{N}} \mathbb{R}^i \rightarrow \textbf{Maybe }\tau \times \mathbb{R} \times \sqcup_{i \in \mathbb{N}} \mathbb{R}^i \rrbracket$:%
\footnote{This definition includes two new types: $\textbf{Maybe }\,\tau$ and $\sqcup_{i \in \mathbb{N}}\mathbb{R}^i$. The $\textbf{Maybe }\,\tau$ type has as elements $\textbf{Just}\, x$, where $x \in \mathbb{\tau}$, and $\textbf{Nothing}$. We take $\textbf{Just}\,x \preceq \textbf{Just}\,y$ if $x \preceq y$, but $\textbf{Nothing}$ and $\textbf{Just}$ values are not comparable. A function $\phi$ is a plot in $\textbf{Maybe }\tau$ if $\phi^{-1}(\{\textbf{Just} \, x \mid x \in \llbracket\tau\rrbracket\})$ and $\phi^{-1}(\{\textbf{Nothing}\})$ are both c-analytic sets, restricted to each of which $\phi$ is a plot. Similarly, for the list type, a function $\phi$ is a plot if, for each length $i \in \mathbb{N}$, the preimage of lists of length $i$ is c-analytic in $U$, and the restriction of $\phi$ to each preimage is a plot in $\mathbb{R}^i$.}
the meaning of a probabilistic program is a function mapping lists of real-valued random samples, called traces, to: (1) the output values they possibly induce in $\llbracket \tau \rrbracket$, (2) a \textit{density} in $[0, \infty)$, and (3) a remainder of the trace, containing any samples not yet used. We can then define $\llbracket\textbf{return}_\tau\rrbracket = \llbracket\lambda x. \lambda\textit{trace}. (\textbf{Just }x, 1.0, \textit{trace})\rrbracket$, $\llbracket \textbf{sample}\rrbracket = \llbracket\lambda \textit{trace}. (\texttt{head}\, \textit{trace}, \texttt{if } (\texttt{length }\textit{trace} > 0)\, (1.0) \, (0.0), \texttt{tail}\, \textit{trace})\rrbracket$, $\llbracket \textbf{score}\rrbracket = \llbracket\lambda w. \lambda \textit{trace}. (\textbf{Just }\, \langle\rangle, w, \textit{trace})\rrbracket$, and $\llbracket\textbf{do }\{x \gets t; s\}\rrbracket = \llbracket\lambda \textit{trace}. \textbf{let }(x_?, w, r) = t\,\textit{trace}\textbf{ in } \texttt{case}_\textbf{Maybe}\, x_?\, (\textbf{Nothing}, 0.0, \textit{trace})\, (\lambda x. \textbf{ let } (y_?, v, u) = (s)\, r\, \textbf{ in } (y_?, w * v, u))\rrbracket$.%
\footnote{Depending on whether its first argument is $\textbf{Nothing}$, $\texttt{case}_\textbf{Maybe}$ either returns the default value passed as the second argument, or calls the third argument on the value wrapped inside the $\textbf{Just}$.}

Let $e : \mathcal{M}\, \tau$ be a closed probabilistic program. Suppose that on all but a Lebesgue-measure-zero set of traces, $\llbracket e\rrbracket(\textit{trace})$ is defined (i.e., not $\bot$\textemdash although the first component of the tuple it returns may be \textbf{Nothing}, e.g. if the input trace does not provide enough randomness to simulate the entire program).\footnote{This condition is implied by almost-sure termination, but is weaker in general. For example, there are probabilistic context-free grammars with infinite expected output lengths, i.e., without almost-sure termination. But considered as deterministic functions of traces (as we do in this section), these grammars halt on all inputs.} On traces where $\llbracket e\rrbracket$ is defined, the following \textit{density function} is also defined: $\llbracket \lambda \textit{trace}. \textbf{let }\, (x_?, w, r) = e\, \textit{trace} \textbf{ in if } (\texttt{length }\, r > 0)\,\,(0.0)\,\,(w)\rrbracket$.  Furthermore, as a function in the language, this density function is $\omega$PAP. Therefore, excepting the measure-zero set on which it is undefined, for each trace length, the density function is PAP in the ordinary sense\textemdash and therefore almost-everywhere differentiable. 
This result was recently proved using an operational semantics argument by \citet{mak2021densities}. The PAP perspective helps reason denotationally about such questions, and validates that AD on trace density functions in PPLs produces a.e.-correct derivatives.

\textbf{Future work.} Besides the ``trace-based'' approach described above, we are working on an extensional monad of measures similar to that of \citet{vakar2019domain}, but in the $\omega$PAP category. This could yield a semantics in which results from measure theory and the differential calculus can be combined, to establish the correctness of AD-powered PPL features like automated involutive MCMC~\citep{cusumano2020automating}. However, more work may be needed to account for the variational inference, which uses gradients of expectations. In our current formulation of the measure monad, the real expectation operator $\mathbb{E}_\mathbb{R} : \mathcal{M\,\tau} \rightarrow (\tau \rightarrow \mathbb{R}) \rightarrow \mathbb{R}$ is not $\omega$PAP, and so we cannot reason in general about when the expectation of an $\omega$PAP function under a probabilistic program will be differentiable.






\bibliographystyle{plainnat}
\bibliography{refs}

\appendix 

\section{Appendix}
\subsection{Type system of the language}
Figure~\ref{fig:type_system} shows the type system of our language from Section 3; all rules are standard.
\input{typesystem}

\subsection{Argument for the soundness of change-of-variables calculations with intensional derivatives}
\label{sec:cov}
Let $\mu$ be a probability distribution over (possibly some sub-manifold of) $\mathbb{R}^n$, with density $\rho$ with respect to a reference measure $B_\mu$. Suppose we have a ``change-of-variables'' algorithm that computes densities $\rho_f$, with respect to a reference measure $B_f$, of pushforwards $f_*\mu$ of $\mu$ by differentiable bijections $f$. The algorithm may use the true derivatives of $f$. We show that such an algorithm also works when $f$ is PAP, if it is given intensional derivatives of $f$ instead of true ones.

Let $f$ be a PAP bijection from $\mathbb{R}^k$ to $\mathbb{R}^m$ (more precisely, it need only be bijective when restricted to the support of $\mu$). Suppose $g$ is an intensional Jacobian of $f$, meaning there exists a (PAP) piecewise representation $\{A_i, f_i\}_{i \in I}$ of $f$ such that $g(x) = J f_i(x)$ when $x \in A_i$. Because the $A_i$ form a partition of $f$'s domain, we have $\mu = \sum_{i \in I} \mathbf{1}_{A_i} \odot \mu$, where $h \odot \mu$ is the measure obtained by scaling $\mu$ by a scalar function $h$. Then $f_*\mu = \sum_{i \in I} f_*(\mathbf{1}_{A_i} \odot \mu) = \sum_{i \in I} (f_i)_*(\mathbf{1}_{A_i} \odot \mu)$. For each $i \in I$, suppose $\rho_{f,i}$ is a density of $(f_i)_*(\mathbf{1}_{A_i} \odot \mu)$ with respect to $B_f$. At points in $A_i$, such densities are soundly computed using the original change-of-variables algorithm, because $f_i$ is differentiable and $g$ gives its (ordinary) Jacobian. Summing these densities (only one of which is non-zero at each point $x$) gives a density $\rho_f$ of $f_*\mu$.

\end{document}

%% file: typesystem.tex
\begin{figure}
\centering
\begin{tabular}{c}
      \\\hline
     $\Gamma \vdash x:\tau$ 
\end{tabular}\,$(x:\tau\in\Gamma)$
\quad
\begin{tabular}{c}
      \\ \hline
      $\Gamma \vdash c:\tau$
\end{tabular}$(c\in\mathcal{C}_{\tau})$
\quad
\begin{tabular}{c}
$\Gamma,x:\tau_1 \vdash e:\tau_2$ \\\hline
$\Gamma \vdash \lambda x:\tau_1.e : \tau_1\to\tau_2$
\end{tabular}
\vspace{.5cm}

\begin{tabular}{c}
     $\Gamma \vdash e_1:\tau_1\to\tau_2$ 
     \quad $\Gamma \vdash e_2:\tau_1$  \\ \hline
     \quad $\Gamma \vdash e_1e_2 :\tau_2$
\end{tabular}
\quad
\begin{tabular}{c}
     $\Gamma \vdash e_1: \mathbb{R}$ 
     \quad  $\Gamma \vdash e_2:\tau$
     \quad  $\Gamma \vdash e_3:\tau$ \\ \hline
     $\Gamma \vdash \texttt{if} \, (e_1 > 0)\, e_2\,e_3 : \tau$
\end{tabular}
\vspace{.5cm}

\begin{tabular}{c}
   $\Gamma,f:\tau_1\to\tau_2 \vdash e:\tau_1\to\tau_2 $ \\ \hline
   $\Gamma \vdash \mu f:\tau_1\to\tau_2. e:\tau_1\to\tau_2$
\end{tabular}

\caption{Type system of our language.}
\label{fig:type_system}
\end{figure}